\documentclass{svjour3}                     % onecolumn (standard format)
\smartqed  % flush right qed marks, e.g. at end of proof
\usepackage{graphicx}
\usepackage{marvosym}
\journalname{Experimental Astronomy}

\begin{document}

\title{ The Astro-WISE datacentric information system}


\author{K. Begeman, A.N. Belikov, D.R. Boxhoorn, E.A. Valentijn}


\institute{ K. Begeman \and  A.N.Belikov (\Letter) \and D. Boxhoorn \and E. Valentijn \at
              Kapteyn Astronomical Institute, University of Groningen, Landleven 12, 9747AB, Groningen, The Netherlands, 
	      \email{\{kgb,belikov,danny,valentyn\}@astro.rug.nl}
	      }

\date{Received: date / Accepted: date}

\maketitle

\begin{abstract}
In this paper we present the various concepts behind the Astro-WISE Information System.
The concepts form a blueprint for  general scientific information systems (WISE) which can satisfy a wide and challenging 
range of  requirements for the data dissemination, storage and processing for various fields in science. 
We review the main features of the information system and its practical implementation. 
\end{abstract}

\section{Introduction}
\label{intro}

Digital astronomical catalogues have been built from the very first moment information technology enabled this, e.g. the first Abell catalogue of clusters of galaxies was 
digitally prepared and printed in the days minus signs were not available in print~\cite{Abell}. As soon as digital scanning devices became available photographic material 
was scanned and first large image surveys were digitally published, such as ESO-LV~\cite{ESO-LV} and the  digitized sky surveys like  DPOSS~\cite{DPOSS}, succeeded 
by the Palomar-Quest Survey~\cite{Quest}, in the 90's followed by CCD-based surveys  such as the 2MASS survey~\cite{2MASS},  the Sloan Digital Sky Survey SDSS~\cite{SDSS}.

The rapid accumulation of astronomical digital data and its public dissemination was, compared to other disciplines, achieved at  an early stage, thanks to an open and collaborative astronomical world community who adopted the FITS image data format as early as 1979~\cite{FITS}.

The CDS, initially Centre de Donn\'ees Stellaires and later renamed into Centre de Donn\'ees astronomiques de Strasbourg took the lead in Europe to collect and disseminate the ever growing 
data sets of a zoo of astronomical observatories and projects. VizieR webservices nowadays provide access to over 9000 catalogues. Numerous astronomical data centers followed,  developing specialized services, for example, the Infrared Processing and Analysis Center and the bibliographical SAO/NASA Astrophysics Data System.

In the early 2000's it was realized that the ever growing data volumes require new approaches: the community becomes the data provider and the International Virtual Observatory and its European branch, the European  Virtual Observatory, developed standards and interfaces to allow individual data centers to publish their catalogs and images in a common framework providing worldwide access to users who can query, cross-match and visualize multiple databases via the internet. A highlight formed the overplotting of data from different experiments, like X-ray satellites and optical ground-based observatories  with some keystrokes on the Aladin interactive software sky atlas.

While the Virtual Observatory focused on the dissemination of published data it was also realized in the early 2000's that the upcoming data deluge required a new approach to the handling of the data stream from the telescope to the science-ready result.  Modern experiments not only involve more data, but also require increasing precision on the various calibrations. The dependency of the final result (sometimes only a few numbers, like cosmological parameter values)  on time-variable  calibrations of very large data sets,   involving the evaluation by  large research teams distributed in smaller groups over various sites sets the basic requirements to the system handling the data. While in classical systems data is delivered in various releases to the public, often re-processing the whole set with higher versions of the code, the  high data rates of modern experiments require an alternative approach, where the up-to-date result is derived on demand by the user.

Thus, we set out to design and implement an integrated {\it datacentric}  system Astro-WISE  in which the processing, storage and administration is integrated in a single environment, providing a {\it living} 
system to both the data producers and the customers. Early reports of this development and implementation have been given in~\cite{V1} and~\cite{V2}. To reach this goal,  traceability of each individual 
data item handled by pipelines or any piece of code is carefully maintained, every data item beyond pixel value is  kept as  Metadata and made persistent and distributed in a relational database with an 
object-oriented view, mapping all the dependencies.

\section{Datacentric Approach to Data Processing}

The datacentric  approach facilitates scientists to cooperate in data analysis and mining by means of the internet, nowadays referred to as e-science. The Astro-WISE information system connects to an own developed 
distributed grid processing system (based on Distributed Processing Unit, DPU), but has been also connected to the EGEE/EGI/BiGGrid-Grid~\cite{JOGC}, particularly for the operations of the Lofar radio telescope. Nowadays, this aspect of 
the system is referred to as cloud computing, but the Astro-WISE datacentric approach at the same time facilitates the sharing and web hosting of all the data handled by the system.

Today, we are just taking the first small steps towards the development of approaches and systems who could handle the upcoming data deluge.  The way of  managing and administration the data will be a key,  
and will require ever improving, refining and self-organizing approaches. The key  notion is that a fruitful approach towards handling the data deluge focuses on the  administration, modeling, standardizing  
and tracking of the data rather than on the processing of the data by CPUs. This  datacentric approach, in which the design of operational systems is driven by  data management and data standard issues 
rather than data processing/cpu/pipelining issues.  Processing systems turn into information systems. First attempts to combine archiving and processing in the 1980s include ESO's  MIDAS table file 
system~\cite{MIDAS}, integrating common tables both accessible in source code and by user interfaces like the MIDAS monitor/prompt. The MIDAS table file system allows both  users and programmers 
to change its content and also registering the modifications in table keywords leading to archives ready for further review and (trend) analysis.   Its development was essential for the production and 
photometric calibration of the images of the ESO-LV galaxies. However, the 600,000 fore- and background objects visible on the images, stars and galaxies, could only be handled with great difficulties 
inside the system, response times running into hours per operations, demonstrating the importance of built-in  scalability.  

This MIDAS Table file system marked a stepping stone to current datacentric approaches which  are enabled by the introduction of modern object-oriented programming languages, data modeling tools 
like UML and XML  and relational databases. 

We have combined these views, systems and requirements in order to design and build  integrated information systems which have the potential for self-organizing and enrichment with new data. Both self-organizing 
and enrichment are processes in time. In fact, the datacentric approach implies the detailed modeling and awareness  of how things change in time. Important changes range from
\begin{itemize}
\item[I.] new data entering the system (ingest) to  
\item[II.]  new or modified source code handling these data.  In turn, the new data entering the system might be 
\item[III.] subject to physical changes (e.g. the gain of a sensor), while source code is often modified on the basis of 
\item[IV.] advancements  of human understanding of the physical changes - our  model of the world also changes (e.g. the cause and modeling of the gain variations of the sensor).  
\end{itemize}
The ideal information system would seamlessly cope with all these changes in time, thus creating a living long term digital preservation environment. Astro-WISE is one of the 
first systems attempting to reach this goal.

For several  years  many  authors predicted the upcoming data avalanche  due to the advancements in digital sensor technology.  Obviously, this has now started, 
both scientific experiments such as LHC and Lofar are running into the tens of Petabytes data acquisition regimes, while text, imaging, genome and internet gathered data collections like e-Bay are 
exceeding similar volumes. This triggered the development of Grid infrastructure that is able to handle Petabytes of data (\cite{EGEE} and~\cite{EGEEData}).

All this has happened in an approximate 20 year time span when the first large digital imaging collection was published, ESO-LV~\cite{ESO-LV}, containing scans of 32,000 galaxy images,   
4 Gbyte of data filling a room full with tapes and populating the very first version of optical media.  

By now, the data avalanche is multidisciplinary and worldwide and involves all aspects of our society, from science to commerce,  public services and health care.  Digital data is gathered, processed, 
distributed and accessed at a continuously growing rate. And this is only the beginning:  in, say, two decades, the same  room full with 4 Gbyte of tapes 20 years ago, and 40 Petabyte of tapes today 
could possibly contain 40 Zetabytes (40 million Petabytes) of data. The growth of the data volume produced by various types of sensor networks met with the growth 
of the disk capacity and search for new methods in the increasing of the memory storage density (see~\cite{MSD}).   
In order to be useful all this data has to be organized, administrated,  managed and distributed.

It took a billion years for cells to develop into extremely complex information systems around the DNA; it has resulted into systems and approaches with an incredible complexity with memory mapping, 
copying, filtering  and distribution mechanisms of data including feedback.  Compared to this, our present data organizing mechanisms in IT are still very simple and rudimentary and we are just making 
the first steps to living self-organizing systems and archives required to deal with and optimally use the ocean of digital data to come. 

However, also in the cell stochastic processes are important, and absence of organization ``from above'' are countered by survival of the fittest type of mechanisms, which appear as self-organizing. 
Also in human endeavored IT,  standards seldom come from above, an exception being  the ASCII character standards, mandated in 1968 by  U.S. President Lyndon B. Johnson to be used in all computers 
purchased by the United States federal government.

{\it I have also approved recommendations of the Secretary of Commerce regarding standards for recording the Standard Code for Information Interchange on magnetic tapes and paper tapes when they 
are used in computer operations. All computers and related equipment configurations brought into the Federal Government inventory on and after July 1, 1969, must have the capability to use 
the Standard Code for Information.}

Interchange and the formats prescribed by the magnetic tape and paper tape standards when these media are used.In practice,  standards are often invented by companies or scientific communities 
which are discriminated in a complex survival of the fittest battle, which involves that many factors that it appears stochastic.

The view,  or better requirement, for future self-organizing information systems and its archives is that it organizes itself for all these changes in time.  It could be seen as a  step towards 
the extremely complex and self-organizing information  systems in a  cell.

The core problem for the definition of an information system is the definition of information itself. The information can be an input data flow from sensor detectors, an archive of transactions, 
trends deduced from the analysis of some data and even results of the data modeling or simulations. As result, information systems will host a variety of subsystems defined by its purpose. 
Examples are a Decision Support System gathering and analyzing trends for a specific business use case (see~\cite{DSS}) or Geographical Information Systems with focus on geospatial data analysis 
and a visualization of the data. 

Nevertheless, it is possible to define primary components of any Information System following simple requirements on storage, processing and sharing data. Such a system should include a data model, 
storage and data processing facilities together with user interfaces. In Astro-WISE we have decided to separate and  store all data beyond pixel/measurement data in a database and all pixel/measurement 
data in files on independent storage media.

This paper does not touch details of the implementation of Astro-WISE information system which are described in other papers but describes the general outline of the Astro-WISE information 
system and specify features which make Astro-WISE unique. For the more in-depth review of particular aspects of Astro-WISE we recommend to read corresponding papers on the optical pipeline implemented in 
Astro-WISE~\cite{pipeline}, quality control realised in Astro-WISE~\cite{quality}, the system of user interfaces and services~\cite{interfaces}, and the hardware solution for Astro-WISE and information 
systems created on the basis of Astro-WISE~\cite{hardware}.

\section{Building a Scientific Information System}
\label{sec:1}    

The development of the Astro-WISE information system, the first one which implemented WISE technology, started from the very practical challenge:  enable a community of researchers distributed over 
the world to process the data of astronomical imaging survey of the OmegaCAM 256 Megapixel camera on ESO's VST telescope at Cerro Paranal (Chile). These scientists should be able to evaluate 
the quality of the data, apply a number of calibrations, share the data within the team and employ distributed resources of a Petabytes scale of data storage and Teraflops capacity in data processing. 

From this our basic requirements on the system are derived:
\begin{itemize}
\item[*] Scalability of the system: any part of the system, i.e., data storage, data processing, metadata management, should be scalable with the increase of incoming data and a number of users 
involved in the data processing. The system should be scalable with respect to the data processing algorithms and pipelines, allowing the implementation of new pipelines and derive improved results from the same raw or intermediate data with new algorithms. 
The scalability of data mining should be possible, i.e. the system should satisfy all possible kinds of requests; from the retrieval of a single data item by identifier to a complicated 
archive study involving multiple complex queries. 
\item[*] Distributed system: any derivation of a result or search for a result should be possible by different users at different sites where the system is implemented.  This makes it possible to optimally use shared resources.
\item[*] Traceability: all activity in the system should leave a clear footprint so that it will be possible to trace the origin of any changes in the data and find an algorithm, program and user 
who created a data item. This allows expert knowledge to be shared amongst all users of the system.
\item[*] Adaptability. The system should be possible to adopt for a number of different scientific use-cases, providing resources, pipelines and expertise to perform a data processing according 
to user's interests on the same data set. 
\end{itemize}

The most general requirements listed above results in a set of more detailed and specific requirements to different components of the information system. First, the ability to share data among 
a number of users becomes more valuable when all these users are working with the same set of standards for the data. Second, the requirement to trace any changes in the data processed by different users with 
different programs and pipelines implies that the system has a common approach to treat the data items on different stages of the data processing and keeps the record of changes in the data. As result 
the basic requirements to the system  dictate the use of a common data model and the common standards for the data storage. Below we will describe the common data model features defined from 
the requirements to the system. 

{\bf Requirements to the Data Model:}
\begin{itemize}
\item[-]  Petabyte systems can not work when anarchy is allowed at the data acquisition.  Users should operate with the same standards for the raw measurements data acquired at the experiment. Moreover, 
this implies a careful  definition of the data taking scenarios, observing templates in astronomy, scan scenarios on sensor systems both for the scientific measurements data and for the calibrations. 
The raw data ingested in the system should be fully and completely described in the system so that users do not need to involve detailed knowledge about the acquired data and can share it with other 
users of the system. {\bf The data model must provide complete data provenance}, i.e., both external data ingested in the system and the data derived inside the system should be described with sufficient level of detail 
to be able to rederive them in the environment external to their original one.    
\item[-] The data model should describe all the data products, on all stages of the data processing  For the user it should be possible to understand the origin of each data item and to trace the data 
item back to the raw data. {\bf The data model must provide  full data lineage}, i.e., each data product and literally each bit of information must be traced back to the origin and creator of the information.  
\item[-] The model should match all scientific use-cases and should allow the flexibility  to modify source code without changing the model as long as this makes sense. Obviously, here boundaries will 
have to be determined by means of inside, thus human, knowledge. This is the most tedious part in the design of the information system, but in practice it turns out that the various processes are well 
defined and dictated by physics. {\bf The data model must be flexible to the changes in the data processing}. 
\item[-] Requirements to trace changes in the data triggered by different versions of the processing software and the necessity to repeat the data processing due to the changes in the pipeline implies 
that all parameters that affect the derived result must be preserved in the system. The user should be able to reproduce the data item  and to find a trend in the data due to the changes in the processing 
parameters or algorithm. At the same time the user should be able to reproduce already existing item to ensure the preservation of the data and software used previously. 
{\bf The data model must allow both data reprocessing and data reproduction}, and the reprocessing should be done with the preservation of the previously produced versions of the data. 
\end{itemize}

When properly defined the data model allows to code, instantiate and trace the data processing for all scientific use-cases. 

Also, the data model should allow to accommodate external data,  as well as provide necessary comparison and checking with these external data sets.  

When the data model is carefully defined our approach is to convert this into a pipeline model, in turn translated into a  hierarchy of Classes. These Classes are then mapped into a database. 
We originally did this with object-oriented databases like  Objectivity, but currently we do this in a relational database and build an object oriented view in order to maintain transparency in the 
class model, and its dependencies.

{\bf Requirements to Infrastructure.} The data model should be {\it shared} among geographically  remote sites of the system, as research groups at various locations are collaboratively adding 
and extracting data from the distributed system. The basic requirement is that in fact any component (database, file system, code base, CPUs) should possibly be present at 
different sites, for a number of reasons: 
\begin{itemize}
\item[-] hardware resources between the sites can be shared, 
\item[-] avoiding single centralized components prevent an overload of resources and duties at that node, 
\item[-] redundancy of the system to single-site malfunctions,
\item[-] sharing of resources between partners allows to reduce the overall cost of the system.     
\end{itemize}
  
For example, the data ingestion to the system can be done in Garching, Germany, the data processing in Groningen, The Netherlands and the image analysis in Napoli, Italy.

{\bf Requirements to the User Interfaces.} The user should be able to trace the data processing, to retrieve data items and identify all its dependencies on other items (mostly calibrations) in the system, 
to initiate  data processing of particular data items and to find a quality estimation for the data. 

As we can see, all these requirements put together are more challenging than requirements on common business information systems. The key difference is that we allow everything, including 
the code base to change in time.

In the following chapter we outline  how our requirements formed  a layout for a scientific information system  capable to share a huge data volume, to process these data and fulfill scientific 
use-cases for a number of users.

\section{WISE Concept}
\label{sec:2}

\begin{figure}
\centering
\includegraphics[width=0.90\textwidth]{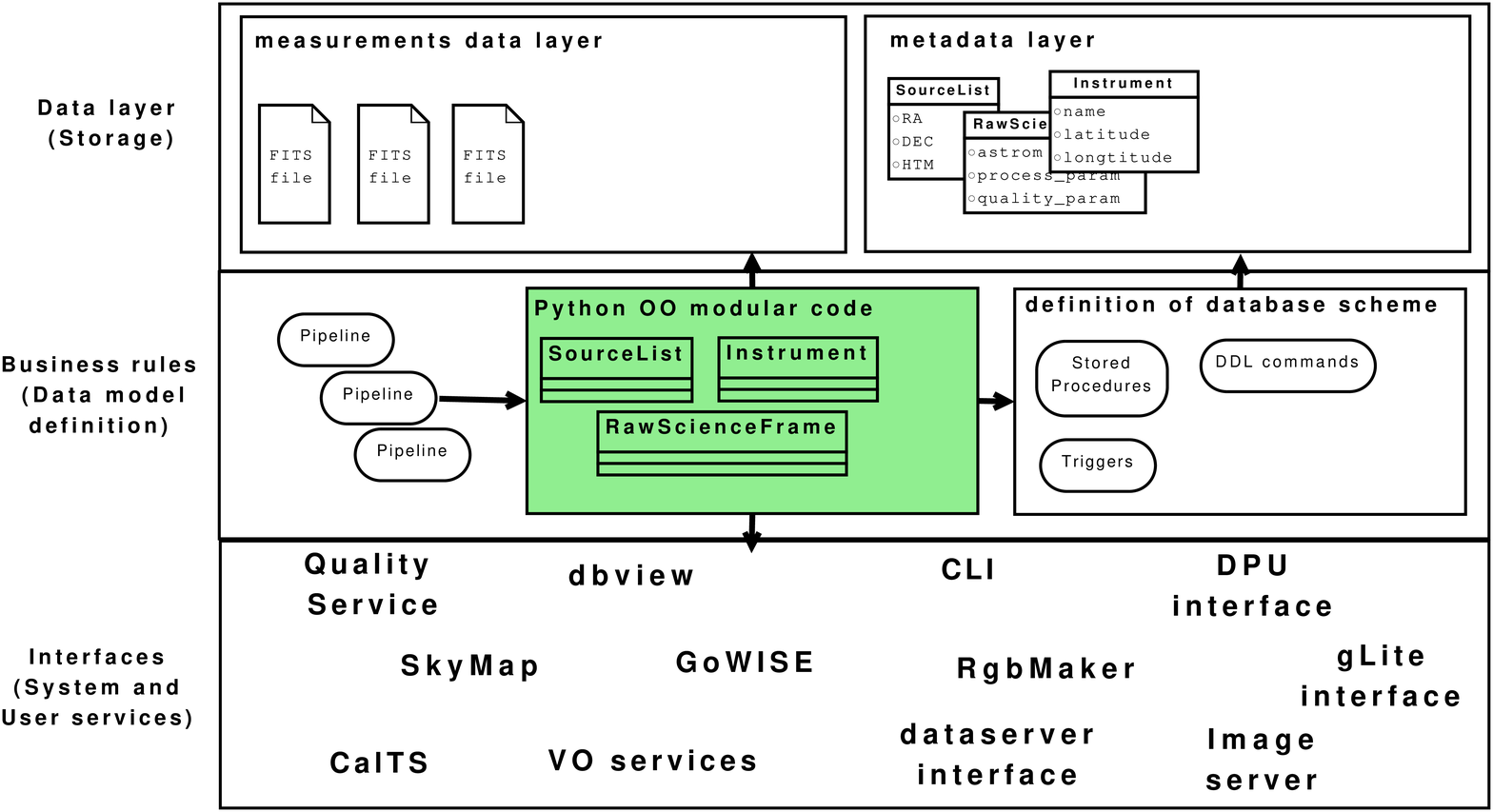}
\caption{Top-level definition of Astro-WISE information system with the specification for each layer.}
\label{fig:AW_definition}
\end{figure}

First of all we have to clearly distinguish between two classes of data storage systems in Astronomy: proper archives containing fixed published data  and information systems. The main 
difference can be found in an ability to work with the data and to change it during the storage. For an archive stored data is a stable entity which is provided to the user as it is, meanwhile 
the information system allows to the user to modify their own  data creating user-specific version of an archive and share this version with other users.

The Astro-WISE approach, and its general multi-discipline WISE approach  implements  a scientific information system. 

The usual approach to building an information system is to perceive  three components: the data layer, business rules and the interfaces (see Fig.~\ref{fig:AW_definition}). In our case, we separate the  data layer in two parts: 
the pure measurement data layer, hereafter data layer, and {\it everything} beyond that, hereafter Metadata layer, ranging from file sizes, to statistics of pixel values and  detected events in 
the measurements (galaxy, particle, word). The Metadata layer allows to implement a data model, the data layer allows to store the data as files in a standard format, while business rules 
implemented in Python classes bind the metadata and the data layers. Interfaces  provide user access  to both the business rules and the data. In the case of Astro-WISE business rules present 
in the Metadata layer (Data Definition Language used to create a data model), data layer (the on-the-fly compression of the data on data storage grid) and--the most apparent to the user---in a number of pipelines and programs which the user defines to process the data. This last part of business rules components we will call data processing layer.

To implement Astro-WISE we use abstractions of storage, processing and database capabilities as a basis for the infrastructure for each of the layers of the system. The metadata layer is realized in a relational DBMS through an abstraction of the required database functionality, the data layer is put on the Astro-WISE 
data storage grid through an abstraction of storage and the processing grid is used to connect the user with the data and metadata layers by a number of interfaces for the data processing layer. Separation of these three 
infrastructures plays a key role in the flexibility of the system, as we will see below.

The metadata layer implements a list of necessary functionalities: 
\begin{itemize}
\item[1.] {\bf Inheritance of data objects}. Using object-oriented programming, all objects within the system can inherit key properties of the parent object. All these properties are made persistent. 
\item[2.] {\bf Full lineage}. The linking (associations or references, or joins) between object instances in the database is maintained completely. Each data item in the system can be traced 
back to its origin. The tracing of the data object can be both forward and backward. For example, it is possible to find which raw frames were used to find magnitudes, shapes and position for 
this particular source and, at the same time, which sources were extracted from that particular raw frame. 
\item[3.] {\bf Consistency}. At each processing step, all processing parameters and the inputs which are used are kept within the system. Astro-WISE keeps the old versions of all data items 
along with all parameters used to produce them and all dependencies between objects. 
\item[4.] Embarrassingly parallel and distributed processing, the administration of asynchronous processing is recorded in the metadata layer in a natural way.
\end{itemize}

Our requirements on distribution and multiple users propagate as key principles of the realization of metadata and  data  layers and business rules which form the core of the WISE approach: 

\begin{itemize}
\item[1.] {\bf Component based software engineering (CBSE)}. This is a modular approach to software development, each module can be developed independently and wrapped in the base language of the system 
(Python) to form a pipeline or workflow.
\item[2.] {\bf An object-oriented common data model used throughout the system}. This means that each module, application and pipeline will deal with the unified data model for the whole cycle of 
data processing from the raw data to the final data product.
\item[3.] {\bf Persistence of all the data model objects}. Each data product in the data processing chain is described as an object of a certain class and saved in the archive of the specific project 
along with the parameters used for data processing.
\end{itemize}

The Astro-WISE system is realized in the Python programming language. It allows to wrap any program into a Python module, library or class. The use of Python also allows to combine the principles 
of modular programming with object-oriented programming, so that each package in the system  can be built and run independently with an object-oriented data model serving as glue between modules. 
At the same time, the logic behind pipelines and workflows in Astro-WISE allows the execution of part of the processing chain independently from the other parts. We will describe this approach 
in more detail in the example of optical image processing in Section~\ref{sec:metadata}. 

The conceptual difference between Astro-WISE and other existing systems is that Astro-WISE moves from the usual for astronomy {\it processing-centric} approach to a {\it data-centric} approach. 
The data processing itself becomes an integral part of the archive. 

The typical solution for the data processing and storage for Astronomy handles the data processing and the final data product delivered to the user as two completely separated entities. 
The data product usually is a result of the processing for the whole survey with fixed processing parameters used for the whole set of images. The  user has access to reduced images and 
the catalog, both data products are stable within a ``release'', which usually refers to the sky coverage performed by the survey. 

In this way the survey data center provides the user with a specific version of the data product that will not change over time, which covers most of the science use-cases.

Nevertheless there are a number of use-cases which can not be satisfied by the ``standard'' version. For example, in the search for objects like brown dwarfs or quasars it is important to lower 
the detection threshold which implies the reprocessing of the data. The user himself has to care about such reprocessing thereby reinventing the whole data processing system for the survey and involving 
his own resources. The Astro-WISE system allows to work on any use-case using the standard pipelines and performing programming on a minimal level - if this is necessary at all. 

Figure~\ref{fig:AW_definition} shows principles of WISE concept in the binding together all layers of the information system. Data processing pipelines define use-cases and data model for the system, which 
is implemented in the Python classes which wrap pipeline and define the common persistent data model objects. The data layer uses this data model, and interfaces are providing an access 
to the system for users. The common data model effectively brings together all layers of the information system.

The common data model can be modified by a mutial agreement of all users of the system. Any user can propose changes in the data model, and, if changes were accepted, the system 
administrator implements these changes in corresponding Python classes and in the database scheme.

In the next sections we will review in detail the infrastructural layers of Astro-WISE and their implementation.

\subsection{Metadata Layer}
\label{sec:metadata}

\begin{figure}
\centering
\includegraphics[width=0.80\textwidth]{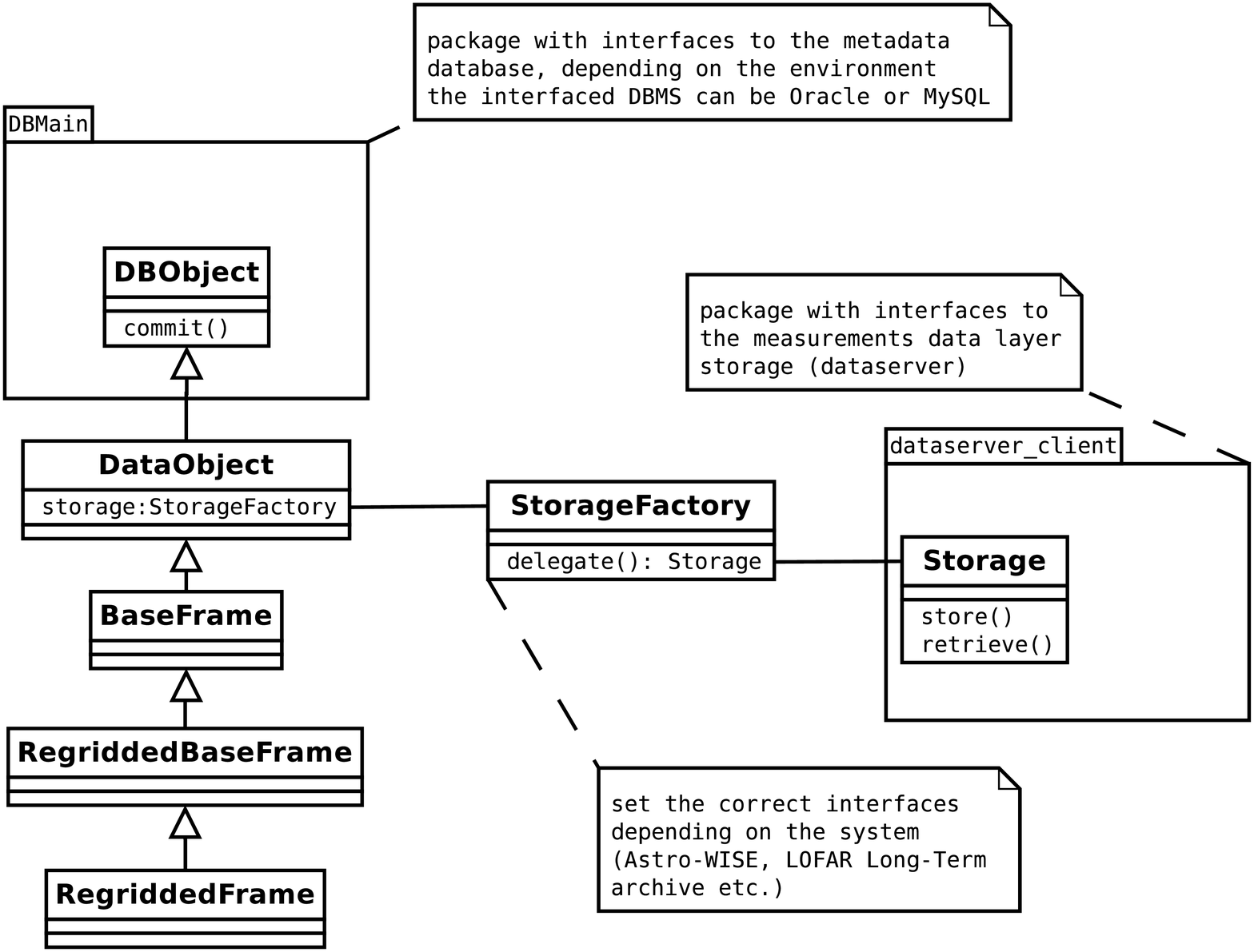}
\caption{Implementation of WISE concept in the data storage. Each persistent Python class inherits interfaces to the metadata and data layers. }
\label{fig:classes}
\end{figure}

\begin{figure}
\centering
\includegraphics[width=0.90\textwidth]{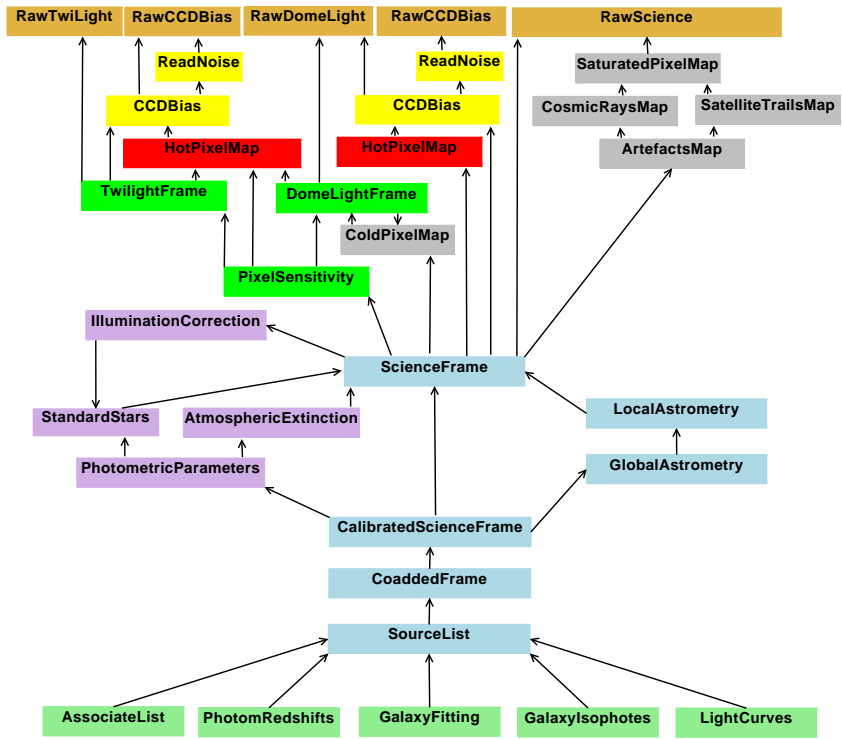}
\caption{A set of Astro-WISE classes for an optical data processing pipeline. Each class is a persistent one stored in the metadata database and linked to the file stored in Astro-WISE system. }
\label{fig:optical_data_model}
\end{figure}

\begin{figure}
\centering
\includegraphics[width=0.90\textwidth]{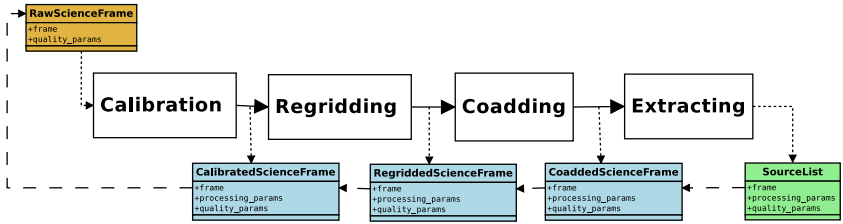}
\caption{Input, intermediate and final data products of the typical processing pipeline.}
\label{fig:optical_process}
\end{figure}

The bulk of the data stored in Astro-WISE is stored in files of some format (FITS). Each file is registered in the 
Astro-WISE metadata database with the unique filename. Apart from just registering each file in the system the metadata 
database implements the important part of WISE approach - the common data model. 

Figure~\ref{fig:classes} shows the general outline for all data models implemented in informations systems based on WISE concept. two core classes 
\verb{DBObject{ and \verb{DataObject{ are parent classes for the class which is storing metadata persistently and the class which is storing the data 
in the file. These classes include interfaces to the metadata database and the data storage, the physical implementation of the metadata database 
and the data (files) storage can be different for different systems. 

Let us suppose that we wish to use Astro-WISE for the data processing of optical images. Figure~\ref{fig:optical_data_model} shows 
typical classes of data items used by the optical data processing pipeline to reduce the data from raw images to the final science ready catalogues. 
This data model, deduced from the pipeline and enhanced keeping in mind possible scientific use-cases, is a central part of the metadata layer. 

The data model is implemented both in the relational database (currently Oracle 11g RAC is used) and in the hierarchy of Python classes. All core classes 
are made persistent, i.e., any change in the object is mirrored in the corresponding tables of the database. The method used for the implementing of the data 
lineage is the Persistent Object Hierarchy (Fig.~\ref{fig:classes}).  According to this method objects of Astro-WISE are made persistent recursively, all operations and attributes of the 
object are saved in the metadata database. 

As an example let us explore the data reduction task - we will create a reduced image from the raw image on one CCD chip. 
First of all, we will select an image we wish to process: 
\begin{verbatim}
awe> raw  = (RawScienceFrame.filename == 'WFI.2000-01-01T08:57:15.410_3.fits')[0]
\end{verbatim}
The image is selected by browsing the metadata database for an object of class \verb{RawScienceFrame{ with filename 
\verb{WFI.2000-01-01T08:57:15.410_3.fits{. The returned object \verb{raw{ was retrieved from the metadata database with 
all attributes. In the next step we will retrieve from the database all necessary calibration objects. The logic in the 
function \verb{select_for_raw{ allows to take the calibration files for the same night of observation (\verb{2000-01-01T08:57:15.410{)
or closest to this night.  
\begin{verbatim}
awe> hot  = HotPixelMap.select_for_raw(raw)
awe> cold = ColdPixelMap.select_for_raw(raw)
awe> flat = MasterFlatFrame.select_for_raw(raw)
awe> bias = BiasFrame.select_for_raw(raw)
\end{verbatim}
Now we can instantiate a new object - our target - a reduced image which will be based on all the images above:
\begin{verbatim}
awe> reduced = ReducedScienceFrame()
awe> reduced.raw  = raw
awe> reduced.hot  = hot
awe> reduced.cold = cold
awe> reduced.bias = bias
awe> reduced.flat = flat
\end{verbatim}
Please, note that all images which will be used to create a new reduced image \verb{reduced{ are referenced through attributes of the 
new image. This new image should get a unique ID, which is a name of the file  where the image will  be stored.  
\begin{verbatim} 
awe> reduced.set_filename()
\end{verbatim}
Now the image can be created invoking the \verb{make(){ method of the \verb{ReducedScienceFrame{ class.
\begin{verbatim}
awe> reduced.make()
\end{verbatim}
The \verb{make(){ produced a file with a new image which will be stored on one of the dataservers of Astro-WISE (see Section~\ref{sec:data_layer}):
\begin{verbatim}
awe> reduced.store()
\end{verbatim}
Finally, the metadata database will be updated and the new image will become a part of the metadata layer. 
\begin{verbatim}
awe> reduced.commit()
\end{verbatim}
The sequence described above is a general way to create a new data entity in Astro-WISE: to combine a set of references to entities which will be used 
to create a new one, to invoke a pipeline or a part of pipeline which will generate a new entity, save the created data in the data layer, commit a new 
data entity to the metadata layer. The processing parameters, for example, the method used for the overscan of the image, will be saved as a persistent attribute of 
the new data entity as well.

The important feature of keeping the full data lineage in the system is an ability to avoid unnecessary reprocessing. The execution of the \verb{make(){ method 
of \verb{reduced{ object actually will start with the search in the metadata database for an object with the same attributes, and if such an object exists (and the user 
is allowed to retrieve it according to the user's permissions) the user will be redirected to an already existing object and the processing part of the method will be skipped. 
The data lineage allows to avoid unnecessary reprocessing as well as to use forward and backward chaining in the dependencies of the data items in the system.  

Objects can be deleted from the database under the following restrictions: every user can delete the data he created at the privileges level 1 (see~\ref{sec:a_a}). 
For higher privileges levels only the project manager is allowed to delete. To delete a data object \verb{myobject{ from the database \verb{Context{ class is used
(see~\cite{pipeline} for the full description):
\begin{verbatim}
awe> Context().delete(myobject)
\end{verbatim}

Objects can be deleted if they are not referenced by other data objects. If they are, it might be desirable to invalidate the object, so that the object will stay in the system 
but will not be used for the data processing (unless the user specifically ask for invalidated objects). 
The user-friendly service for the validation of data objects is described in~\cite{quality}.

\subsection{Data Layer}
\label{sec:data_layer}

\begin{figure}
\centering
\includegraphics[width=0.80\textwidth]{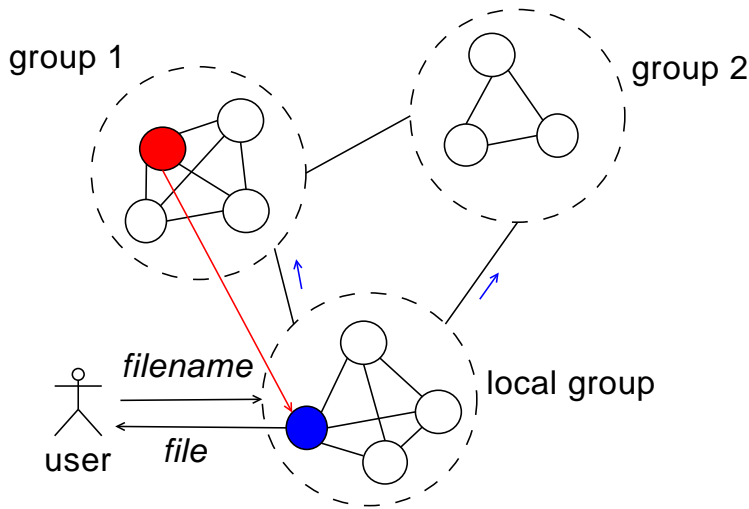}
\caption{Data storage network of Astro-WISE. Dataservers are grouped geographically with one dataserver dedicated to the external data exchange. The user requests a file from a local dataserver and, if the file is not found in the local group, the local dataserver will request the file from the other dataserver groups. As soon as file is found it is copied to the dataserver that the user contacted and provided to the user. }
\label{fig:dataservers}
\end{figure}

Each data entity in Astro-WISE has two parts: a data part which is stored in a file and a metadata which is a part of the metadata database and which keeps all dependencies of the data entity. 
The data part of the entity is stored as a file on one of the Astro-WISE dataservers. The predefined file format can be changed, but for most files the format is FITS. Nevertheless 
the dataserver is not limited to a single file format and can store files of any type. The dataserver is a specific solution which in its functionality is closests to the storage 
element of the Grid.

The main requirement to the dataservers as storage space is, apart from the safety, scalable size. The size should be scalable at the Terabyte level allowing to increase 
the storage volume from a few Terabytes to almost a Petabyte (this is the typical scale for the KIDS data archive  from a few raw images at the 
beginning of the survey to hundreds of Terabytes with all intermediate data products at the end of observations).

Dataservers are organized in geographically close clusters, within each cluster all dataservers know about the existence of all other dataservers in the cluster. A dataserver in one cluster 
can contact other clusters via a dedicated dataserver in another cluster (see Fig.~\ref{fig:dataservers}). All communications are done using a standard HTTP protocol. Two types of requests 
are available: to retrieve a file and to store a new file.

The unique name of the file is used as a unique identifier and this unique name is stored in the database as part of the metadata.

The user of the system does not know the actual location of the file on the underlying file system and operates with the URL of the file only. The URL has the form \verb{http://<data server address>/<file name>{. Each Astro-WISE service has pre-defined dataservers which are used retrieve data. Usually the administrator of the local Astro-WISE node assigns this dataserver selecting ``geographycally closest'' dataserver. The user can change this assignment and use the dataserver the user prefers. 

On the request from the user to retrieve the file the dataserver can then either return the requested file or redirect the client to the dataserver that has the file. Each dataserver has a permanent data storage and a cache, the last one is used to temporarily store a file retrieved from the other dataserver. By  user request, for example, 
\verb{http://ds.astro.rug.astro-wise.org:8000/WFI.2000-09-28T02:22:37.466_8.fits{ the dataserver \verb{ds.astro.rug.astro-wise.org{ will check cache space for the file 
\verb{WFI.2000-09-28T02:22:37.466_8.fits{. If the file is not found, the dataserver will check it's own permanent storage space, and if there is no such file, all dataservers in a 
``local'' cluster (\verb{astro.rug.astro-wise.org{) will be requested. If there is still no file found, all other clusters will be requested, and as soon as the file is located it will 
be copied to the cache of the \verb{ds.astro.rug.astro-wise.org{ and returned to the user. 
In addition, 
a file can be compressed or decompressed on-the-fly during retrieval, and a slice of the data can be
retrieved by specifying the slice as part of the URL.

The dataserver is written in Python and can be installed on any operating system and underlying filesystem that can 
support long case-sensitive filenames. For the Astro-WISE Linux filesystems, XFS and GPFS are currently in use.

\subsection{Processing Layer}
\label{sec:processing}

The example in the Section~\ref{sec:metadata} involves the processing facilities of Astro-WISE, which are distributed and combined from processing facilities of all Astro-WISE 
partners. The user has the choice to send the job to one of the processing elements of Astro-WISE (including Grid computing elements) or to use the processing power at the user's disposal, 
for example, PC or even notebook. 

In the core of the processing layer of Astro-WISE is a distributed processing unit (DPU). This is a three-component middleware which consists of the DPU server, DPU client and the
so called DPU runner. The DPU server is a front-end to any processing system and interacts with the processing system's native queuing system, allowing the user to submit jobs 
to this particular queuing system, inspect or cancel them. The DPU client includes all functions and methods the user can call to interact with DPU server. The DPU runner is a 
program which is run on the remote processing facility and checks availability of all necessary software to run an Astro-WISE job, installing required packages if necessary. 
The system runs on {\it openpbs} or under its own queue management software. The DPU itself allows synchronizations of jobs as well and can also transfer parts of a
sequence of jobs to other DPU's.

In the case of Grid computing element the DPU server will check a user's identity and will use the user's credentials (Grid certificate) to mediate with the Virtual Organization Management System. Currently OmegaCEN is using {\it omegac} Virtual Organization with an ability to submit jobs to Grid computing elements in Amsterdam and Groningen.

\subsection{Interfaces}

The description of Astro-WISE system would be incomplete without the description of a number of interfaces provided for the user. The user can 
write his own applications in the Python language calling Astro-WISE libraries or can involve Astro-WISE services. The first case requires the use 
of the Astro-WISE Command Line Interface called AWE (Astro-WISE Environment), which can be installed on any site, PC or notebook

The Command Line Interface - CLI - supposes that the user writes his own  programs using Python, but to browse the data or even to process observations there is no need to use the CLI. All operations 
required to perform this activity are possible with a set of standard web services of Astro-WISE. Of course, the use of the CLI gives to the user much more freedom in data 
processing. The CLI is more useful for experienced users working on a particular use-case, meanwhile web services are developed for routine operations during the data processing 
of the surveys.

The web interfaces are divided into two types: data browsing/exploration and data processing/qualification. 
The first group includes:
\begin{itemize}
\item[*] dbviewer\footnote{http://dbview.astro-wise.org/} -- the metadata database interface which allows browsing and querying of all attributes of all persistent classes stored in the system,
\item[*] quick data search\footnote{http://gowise.astro-wise.org/} -- allows querying on a limited subset of attributes of the data model (coordinate range and object name), and provides results of all projects in the database,
\item[*] image cut out service\footnote{http://cutout.astro-wise.org/} and color image maker\footnote{http://rgb.astro-wise.org/} -- these two services are for 
the astronomical image data type and allow to create a cut out of the image or to create a pseudo-color RGB image from three 
different images of the same part of the sky,
\item[*] GMap\footnote{http://skymap.test.astro-wise.org/} -- exploration tool of the Astro-WISE system using the GoogleSky interface.
\end{itemize}
Data processing / qualification interfaces are:
\begin{itemize}
\item[*] target processing\footnote{http://process.astro-wise.org/} -- the main web tool to process the data in Astro-WISE. 
This web interface allows users to go through pre-defined 
processing chains, submitting jobs on the Astro-WISE computing resources with the ability to select the computing node of Astro-WISE,
\item[*] quality service\footnote{http://quality.astro-wise.org/} --  allows to estimate the quality of the data processing and set a flag highlighting the quality of the data,
\item[*] CalTS\footnote{http://calts.astro-wise.org/} -- web interface for identifying and qualifying calibration data. 
\end{itemize}

All web services are built using a set of Python classes developed for Astro-WISE as a basis and uses a modular principle, which allows to create a new web service 
using components of older ones.

\subsection{Authorization and Authentication System}
\label{sec:a_a}

Astro-WISE is a multi-user system which must accommodate sharing data between scientists and at the same time protect the private data of each user. Each user in the Astro-WISE system has 
an identity protected by a password, and, optionally, if the user wants to submit job to Grid resources he has to get a Grid certificate.

The authorization and authentication system is implemented on the level of the metadata database. As soon as a user logged in with his username/password, the user's privileges are 
checked in the database, allowing the user to browse the data according to the user's privileges while obeying the privileges of other users. The data in the metadata database are grouped by {\it projects}. A Project is a 
collection of resources which is associated with a group of users who can access these resources, usually one of these users has the role of {\it Project Manager} who has privileges to include 
new users, remove users and publish the data to the wider community.

The system of access to the data is based on three attributes which any data entity in Astro-WISE has: {\it user}, {\it project} and {\it privileges}. The first one identifies the user that created the data entity.
The second one defines to which project 
the data entity belongs, the third one defines who is able to use this data entity. All these attributes are initialized the first time the data entity is made persistent in the Astro-WISE system 
(including the case that the entity is created by one of the Astro-WISE pipelines) and they are both persistent attributes, i.e., stored in the metadata database.

Table~\ref{table:privileges} shows the range of values for the {\it privileges} attribute. In the case of {\it privileges=1} only the creator of the data item can see it, the creator can 
raise {\it privileges} to {\it privileges=2}, in this case all the users of the project will be able to browse this data entity. Raising {\it privileges} to 3 the user makes the data item 
accessible for all Astro-WISE users in all projects, and with {\it priveleges=4} the anonymous user (a user with minimal read-only privileges in the system) can see it as well. Finally, 
with {\it priveleges=5} the data item is accessible via Astro-WISE Virtual Observatory interfaces. 

\begin{table}
\caption{Privileges system of Astro-WISE}
\label{table:privileges}
\begin{tabular}{ | l | c | c | c | c | c | c | }
\hline
privileges       & scope    & visible to      \\
                 &          &                 \\
\hline
1             & private     & the owner only         \\
2             & project     & users of the project         \\
3             & Astro-WISE  & authorized users of Astro-WISE        \\
4             & world       & anonymous users of Astro-WISE        \\
5             & VO          & published in Virtual Observatory         \\
\hline
\end{tabular}
\end{table}

The special Python class {\it Context} was created to handle authorization and authentication. The {\it Context} allows to change privileges 
of all user's data items, during the change {\it Context} is checking for dependencies to prevent inconsistency in the dependencies due to 
the different privileges of data items. For example, if the raw image was published by the user to the project scope and some other user has created 
a reduced image from this raw image, the original user will not be able to downgrade privileges on the raw image to the private data scope.

\subsection{WISE architecture}
\label{sec:architecture}

\begin{figure}
\centering
\includegraphics[width=0.90\textwidth]{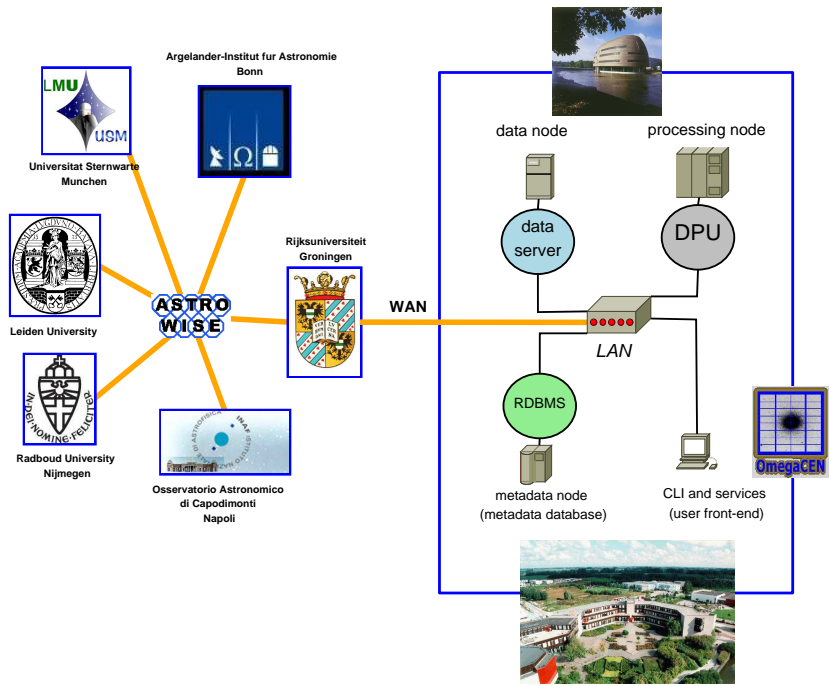}
\caption{Present-day Astro-WISE nodes with a typical composition of a node in a full deployment node.}
\label{fig:AW_node}
\end{figure}

We described above three infrastructural components of Astro-WISE: the relational DBMS (the metadata layer), dataserver (the data layer) and 
DPU (the processing layer). Combined with user interfaces these components build an Astro-WISE node - a detached Astro-WISE site which can operate 
independently from other sites. Multiple nodes can be combined to form a distributed system. Each node is independent from others in the sense that it is administrated independently and can 
handle both the data distributed over other nodes and the data restricted to this node only and shielded from other nodes. 

Figure~\ref{fig:AW_node} shows the elements of the node:  
\begin{itemize}
\item[*] Data storage servers (dataservers).
\item[*] A metadata database to store a full description of each data file with links and references to other objects. 
\item[*] Astro-WISE programming environment (CLI) along to web services which give the user an 
ability to access to the stored data and launch a data processing. 
\item[*] Computing nodes for the data processing with DPU servers on top of their quing system. 
\item[*] A version control system for developers to include new modules, classes and libraries into the system. 
\end{itemize}
All these elements are optional, an Astro-WISE node can be installed without dataservers (dataservers from other nodes are used), 
metadata database or processing facilities. In fact an Astro-WISE node can be installed on a notebook as Astro-WISE environment only giving the user access 
to the system - if the user is not satisfied with web services on the remote nodes. 

Presently Astro-WISE includes sites at Groningen University, Leiden University and 
Nijmegen University (The Netherlands), Argelander-Institut f\"ur Astronomie, Bonn and  Universit\"ats-Sternwarte M\"unchen
(Germany) and  Osservatorio Astronomico di Capodimonte, Napoli (Italy).

\section{Astro-WISE: migration to other systems}

The WISE concept for an information system is a set of principles described in Section~\ref{sec:2}, Astro-WISE is the first system which realizes this concept and serves as a basis 
for the further development of information systems. In this section we describe the adaptivity of the parent system to new tasks and challenges on the example of LOFAR Long Term Archive 
and Molgenis system. 

The importance of the LOFAR Long Term Archive (LTA) for the development of WISE concept is a necessity to use an external infrastructure which should be included in the architecture to form 
a complete information system. We preserved all principles of Astro-WISE adding storage and processing which is not controlled by the system itself (BiGGrid\footnote{http://www.biggrid.nl/}). 
Additionally we integrated three different systems of Authorization and Authentication to make it possible for the LOFAR observatory to create new users and control resources in the system. 
The LOFAR LTA design is described in~\cite{LTA_design} and the architecture and  infrastructure in~\cite{LTA_architecture}.  
 
Another significant development was achieved with the adaptation of Molgenis\footnote{http://www.molgenis.org}. Molgenis is a framework written in Java to build user interfaces and databases 
from definitions that are written in XML. From these definitions the database tables and webserver are generated. Its origins lie in biomedical applications. Historically Molgenis focused more on 
interaction with the user to streamline the ``protocols'' (recipes for analysis) of researchers to keep track of analyses of results. On the other hand, the WISE technology focused more on results from a 
processing perspective. This lead to the idea to combine the strengths of both frameworks and use some models that already existed in XML for Molgenis to generate a datamodel in Python for Astro-Wise. 
Then Molgenis could be extended to use the WISE infrastructure for distributed storage and processing.

The existing Molgenis XML model is organized in a common fashion through ``module'', ``entity'' and ``field'' where each field has a type of {\it int}, {\it float}, 
{\it str} or can be a reference type such as {\it mref} or {\it xref}. Most of these have a counterparts in the WISE framework which made it possible to write a conversion tool. 

The webserver that is generated by Molgenis has to communicate to a database that it usually creates itself. However, in this case the database had to be generated by Astro-Wise because 
the frameworks did not have a backend for a common database (Oracle vs. Non-Oracle). Instead of writing a new database backend for either framework it was decided to write an xmlrpc interface 
to encapsulate database queries and return their results. Since client software for all database flavours does not have to be present this functionality can be extended, e.g., since this is xmlrpc, 
a programming language independent implementation could also use the database in a way that the WISE framework dictates. 

The approach developed and tested in the case of Molgenis allows to create a new information system based on Astro-WISE with all services starting from the data model coded in XML and to do this 
automatically. This approach allowed to decrease the time and resources spent for the developing and implementation of a new system significantly. 

The next system which will be created with the approach tested on Molgenis is a data processing system for Multi Unit Spectroscopic 
Explorer\footnote{http://www.eso.org/sci/facilities/develop/instruments/muse} (MUSE).

\section{Conclusion and Future Work}

The Astro-WISE information system, the first information system in Astronomy, proved to be a reliable and flexible tool for the data processing. Originally developed to process the data of KIlo 
Degree Survey (KIDS\footnote{http://www.astro-wise.org/projects/KIDS/}) it triggered development of the unique approach to the architecture of scientific information systems (WISE approach).

Both Astro-WISE and the WISE approach are living systems which are open to improvements. For the last 2 years further development of the WISE approach is hosted by Target Holding\footnote{http://www.rug.nl/target}. 
Target Holding is an expertise center in the Northern Netherlands which is building a cluster of sensor network information systems and provides cooperation between a number of scientific projects and 
business partners like IBM and Oracle. Target creates and supports a hardware infrastructure for hosting tens of Petabytes of data for projects in astronomy, medicine, artificial intellegence and biology. 

In this development the WISE approach was used to create new information systems extending the original Astro-WISE on new data models, new data storage and processing capacities and new fields.

\begin{acknowledgements}
Astro-WISE is an on-going project which started from a FP5 RTD programme funded by the EC Action ``Enhancing Access to Research Infrastructures''. This work was performed as part of the Target project. 
Target project is supported by Samenwerkingsverband Noord Nederland. It operates under the auspices of Sensor Universe. It is also financially supported by the European fund for Regional Development and 
the Dutch Ministry of Economic Affairs, Pieken in de Delta, the Province of Groningen and the Province of Drenthe. 
\end{acknowledgements}



\end{document}